\documentclass[runningheads]{llncs}

 \usepackage{graphicx}
\usepackage{hyperref}
\usepackage{caption}
\usepackage{subcaption}
\usepackage{tabularx}
\usepackage{color}
\usepackage{color}
 \usepackage{amssymb,mathtools}
 
 \usepackage[nocompress]{cite}

\author{Md Iqbal Hossain and Stephen Kobourov}

\institute {Department of Computer Science\\ University of Arizona}

\titlerunning{Research Topics Map, {\tt rtopmap}}
\begin{document}
	
	\title{Research Topics Map: {\tt rtopmap}}
 	\pagenumbering{arabic}
	\maketitle 
		 
	\vspace{-0.4cm}\begin{abstract}
 In this paper we describe a system for visualizing and analyzing worldwide research topics, {\tt rtopmap}. We gather data from google scholar academic research profiles, putting together a  weighted topics graph, consisting of over 35,000 nodes and 646,000 edges.  The nodes correspond to self-reported research topics, and edges correspond to co-occurring topics in google scholar profiles. The {\tt rtopmap} system supports zooming/panning/searching and other google-maps-based interactive features. With the help of map overlays, we also visualize the strengths and weaknesses of different academic institutions in terms of human resources (e.g., number of researchers in different areas), as well as scholarly output (e.g., citation counts in different areas). Finally, we also visualize what parts of the map are associated with different academic departments, or with specific documents (such as research papers, or calls for proposals). The system itself is available at \url{http://rtopmap.arl.arizona.edu/}.

	\end{abstract}

    \section{Introduction}\vspace{-.3cm}\label{se:introduction}
\label{Introduction}

\begin{figure}[t]
	\hfil \includegraphics[width=1\textwidth]{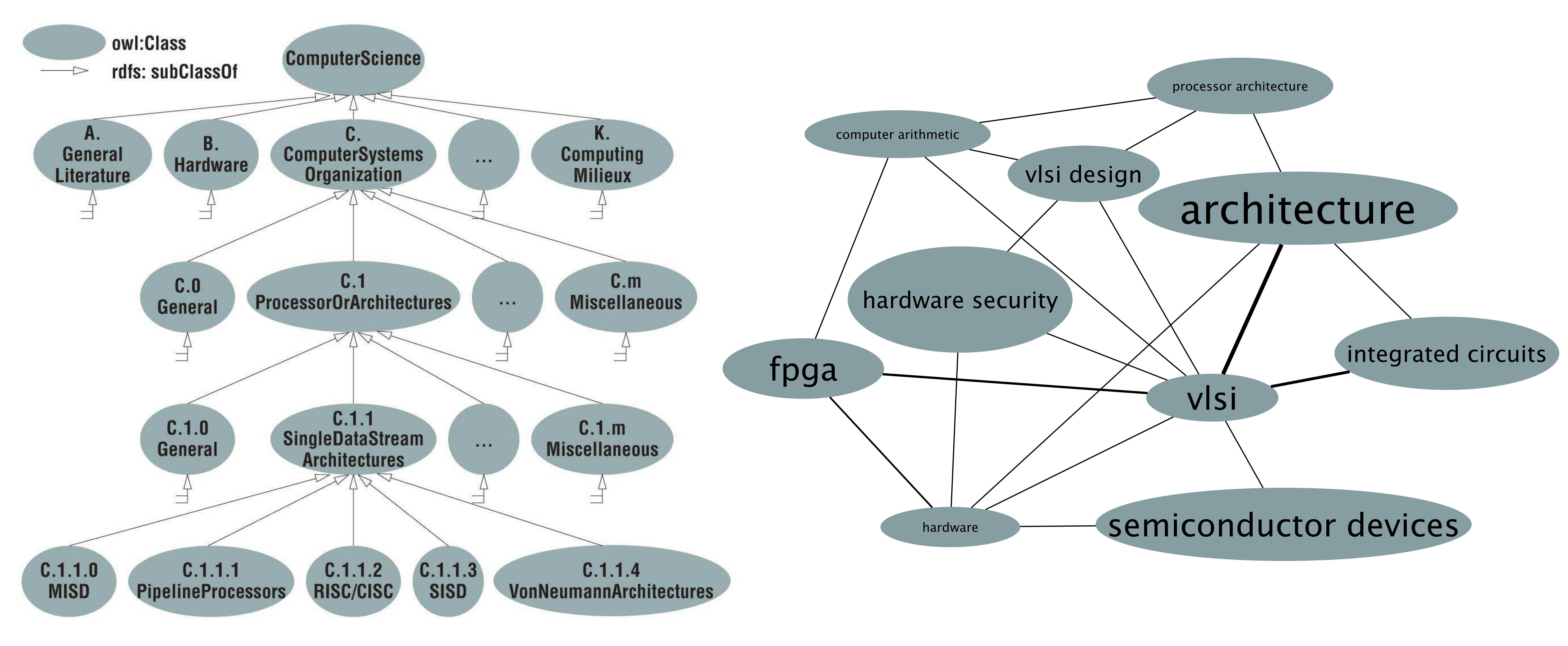} \hfil
	\vspace{-.3cm}\caption{Part of the ACM classification (left) and a more realistic network showing different types of connections between topics (right).\vspace{-.5cm}}
	\label{fig:topicsnetwork}
\end{figure}

Cataloguing and organizing science often involves taxonomies, ontologies, and knowledge graphs, but most often research topics are categorized in hierarchical trees~\cite{amsClassification,Effendy17}; see Fig.~\ref{fig:topicsnetwork}. For example, ``Hardware" and  ``computer systems organization" are subfields of ``computer science."
Knowledge graphs make it possible to see more of the connections between topics than can be embedded in a tree, however, the ability to show clearly the underlying hierarchical structures is compromised. 

Maps have guided human exploration for many centuries and recently there have been several efforts to visualize scholarly knowledge and research expertise using topic maps. Basemaps of science can be generated, for example, by analyzing citation links between publications and placing similar records next to each other.
Such maps can be used to compare expertise profiles, to understand career trajectories, and to communicate emerging areas, as illustrated in the special {\it PNAS} issue on ``Mapping Knowledge Domains"~\cite{mapkd}, and B\"orner's ``Atlas of Science"~\cite{a1} and ``Atlas of Knowledge"~\cite{a2}.

Most maps of science are really node-link diagrams with one level of detail; a few support two or more levels (e.g., the UCSD map of science which is the current standard 
 has two levels of detail)~\cite{borner_design_2012}. However, people have difficulty reading large-scale networks~\cite{is-sensem} and few can derive knowledge from multi-level representations of networks. 
Given encouraging results about the effectiveness of map-like visualization of large graphs~\cite{nlg,saket2015map}, we adopt the Graph-to-Map (GMap) framework~\cite{Gansner10} to visualize and explore our research topics map. 
The GMap visualization of relational data was introduced in the context of visualizing recommendations,
where the underlying data is TV shows and the similarity between
them~\cite{Gansner09} and has already been used to visualize research topics in computer science publications~\cite{Fried14}.

\begin{figure}
	\hfil \vspace{-.3cm}\includegraphics[width=1\textwidth]{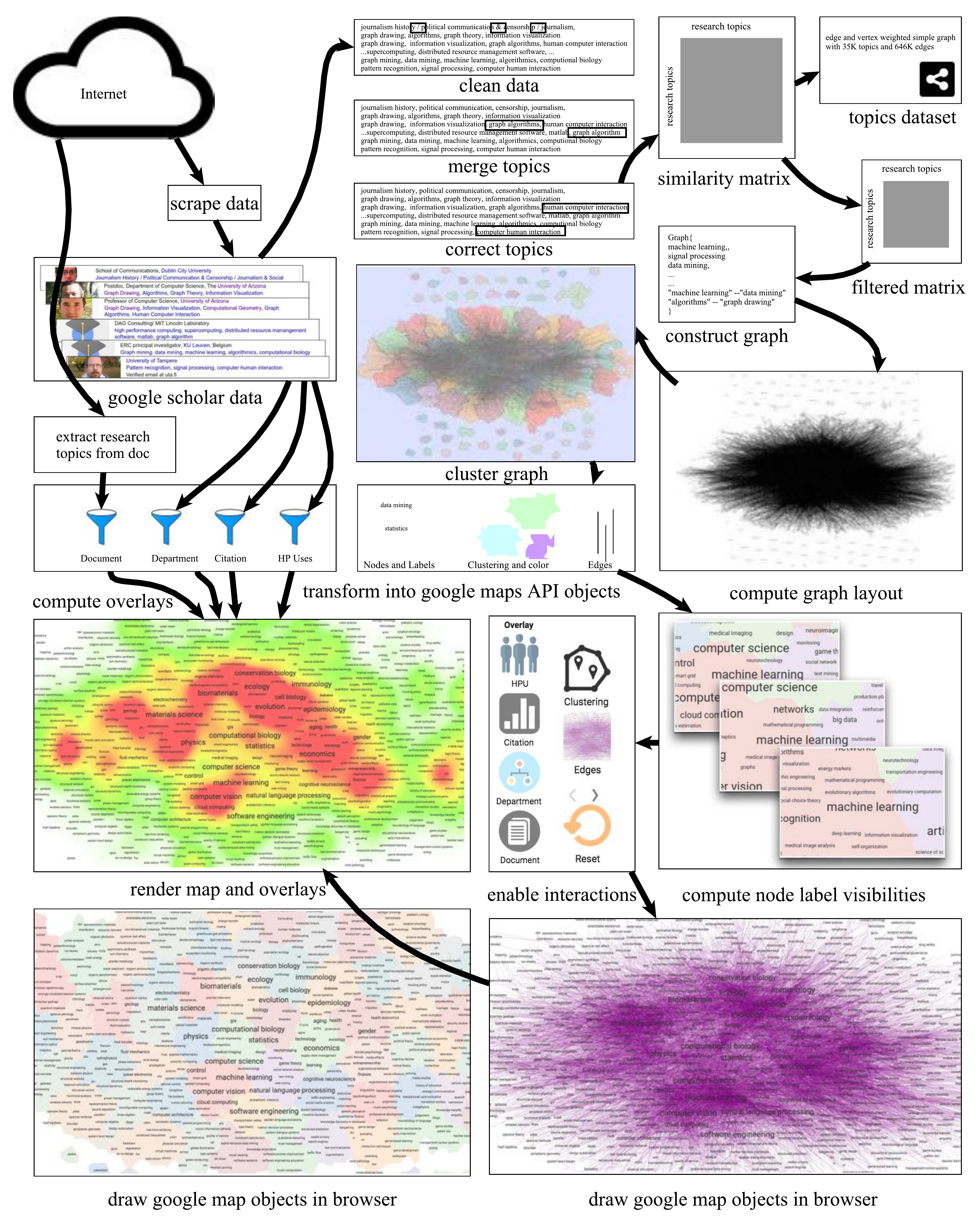} \hfil
	\vspace{-.35cm}\caption{Overview of the {\tt rtopmap} system.}\vspace{-.6cm}
	\label{fig:overview}
\end{figure}

Our research topics map system, {\tt rtopmap}, covers all research topics indexed by google scholar and provides the ability to show 
 human resource investments (e.g., number of researchers in different areas) and scholarly output (e.g., citation counts in different areas) of different universities. 
 The system supports zooming/panning/searching and other google-maps-based interactive features, including several map overlays showing what parts of the map are associated with different academic departments, or with specific documents; see 
Fig.~\ref{fig:overview} for an overview of the system.
  
We gather data from google scholar and then 
 clean, split, merge, and correct the research topics (which become the nodes in the graph). We next compute a similarity matrix based on co-occurrence of topics in scholar profiles, which is used to place edges between topics that are frequently listed together. This gives us the topic network. We reduce the size of the graph by removing rarely occurring topics and weak connections. We then use a multi-level force-directed placement, node overlap removal, and clustering algorithms to represent the graph as a map. 
Nodes, node labels, polygon colors, and edges are transformed into google map objects, which are then 
drawn in the browser using the google maps API. Eight different level-of-detail (zoom levels) are precomputed, determining which nodes are present on a given level, computing label font sizes, and ensuring no label overlaps.
Different overlays are added on demand.

\section{Related work}
\vspace{-.3cm}

Today, the most comprehensive map of science and classification uses ten years of paper-level data from Thomson Reuters' Web of Science and Elsevier's Scopus to group about 25,000 journals into 554 subdisciplines that are further aggregated into 13 disciplines; see data and detailed procedure in~\cite{BKP12}. However, the two-level map of 13 disciplines and 554 subdisciplines is too coarse for organizing, navigating, managing, and making sense of millions of publications. 

Microsoft's Academic Graph database has 50,000 fields of study (FOS)~\cite{sinha2015overview}. Three levels of relationships are present among the fields, although field importance is not measured or quantified. A FOS score based on researcher and citation counts has been proposed for computer science~\cite{Effendy17}.
Hug {\it et al.} analyze FOS and report that they tend to be dynamic and too specific, while field hierarchies are incoherent~\cite{HugOB16}.
Liu {\em et al.}~\cite{liu2014hierarchical} use a hierarchical latent tree model (HLTM) to extract a set of hierarchical topics to summarize a corpus at different levels of abstraction. In HLTM, a topic is determined by identifying words that appear with high frequency in the
topic and with low frequency in other topics. Yang {\em et al.}~\cite{Yang2017Qu} use a HLTM in their visual analytics system, VISTopic.
Mane and B\"orner~\cite{Mane04} visualize 50 frequent and bursty words in their analysis of publication of the Proceeding of the National Academy of Sciences.

Words from paper titles have also been used as indicators
for the content of a research topic, and visualizations based on this approach have been studied~\cite{van2006mapping,Fried14,Zhou2016}. 
  Many earlier approaches focus on analyzing specific journals, conferences, or research areas, e.g., analyzing computer science conferences and journals~\cite{Fried14}, trends in computer science research~\cite{Effendy17}, the International Conference on Data Mining (ICDM)~\cite{Misue14}, publications in data visualization~\cite{Heimerl16}. 
Domenico {\it et al.}~\cite{DeDomenico2016} quantify attractive topics (i.e., topics that attract researchers from different areas). 
Sun {\it et al.}~\cite{Sun16} build a network, with computer science conferences as nodes and edges between two conferences with common authors. 
  Map-based visualization has been used for document visualization~\cite{Wise95,skupin2002cartographic}.

Citations are considered an important contribution measurement~\cite{zhao2015analysis} and are used in visualizations of scholar profiles~\cite{Portenoy2017} and paper recommendation systems~\cite{West16}. 
Citation analysis with data from the Web of Science~\cite{perianes2015university} and from Microsoft's Academic Graph~\cite{HugOB16} have been considered.
CiteRivers~\cite{Heimerl16} and CiteVIS~\cite{stasko2013citevis} analyze and visualize IEEE VIS conference citations, as do Ke {\em et al.}~\cite{ke2004major}. 


 
 

 

  

Also related to our work are many of the graph visualization techniques and tools. 
Graph layout algorithms are also provided in several libraries, such as GraphViz~\cite{graphviz}, OGDF~\cite{chimani2011}, MSAGL~\cite{nachmanson2008}, and VTK~\cite{schroeder2000}, which however, do not support interaction, navigation, and data manipulation. Visualization toolkits such as Prefuse~\cite{heer2005}, Tulip~\cite{auber2012}, Gephi~\cite{bastian2009}, and yEd~\cite{yed} support visual graph manipulation, and while they can handle large graphs,
their rendering does not: even for graphs with a few thousand vertices, the amount of information rendered statically on the screen makes the visualization difficult to use.

There are research papers that describe interactive multi-level interfaces for exploring large graphs such as ASK-GraphView~\cite{abello2006}, topological fisheye views~\cite{GKN05}, and Grokker~\cite{rivadeneira21}. Software applications such as Pajek~\cite{de2011} for social networks, and Cytoscape~\cite{shannon2003} for biological data provide limited support for multi-level network visualization. These approaches rely on meta-graphs, made out of meta-vertices and meta-edges, which make interactions such as semantic zooming, searching, and navigation counter-intuitive. Not many of the tools and systems above provide browser-level navigation and interaction for large graphs.



Our work differs from earlier related work in several important ways: (1) we collect the underlying data using a bottom-up approach, based on self-reported data from the actual researchers, rather than using top-down taxonomies and ontologies, (2) our visualization provides map functionality (multiple zoom levels, searching, overlays), and (3) the ability to customize both the underlying base map and the overlays.

	\vspace{-.2cm} 
\section{Network Generation}\vspace{-.3cm}
The set of research topics is not fixed or even well defined, as new topics are continuously created while old ones fade away. Automatically extracting keyword based topics from the research literature is a popular approach~\cite{Yang2017Qu,Fried14}, but has many limitations, such as identifying general topics (e.g., mathematics, physics) and specific sub-topics (e.g., graph drawing, network visualization). 
We use self-reported research topics from google scholar. 
Before we can build the topic graph, we scrape the data and then
 clean, split, merge, and correct the research topics.
 Next we build a similarity matrix $M$ with topics as rows and columns. The value $M(i,j)$ represents the similarity between the pair of topics $(i,j)$,  based on co-occurrence of the two topics in scholar profiles. 
The complete network is quite large, containing about 35,000 topics and 646,000 edges.

We reduce the size of the network by removing nodes and edges with low weights. Node weight is directly proportional to the number of scholar profiles that contain that topic, and edge weights are directly proportional to the number of scholar profiles that contain both topics.
We remove a large number of infrequent topics, topics that contain typos, and topics listed in languages other than English. 

\medskip \noindent{\bf Data Scraping:} 
While some analysis of google scholar data exists~\cite{jacso2005google,falagas2008comparison,bar2007h}, there is not much work based on data extracted from google scholar. 
Data retrieval is laborious due to the lack of API and metadata scarcity~\cite{bornmann2016application}. 
We started with a list of 1,000 universities~\cite{top1000} and then requested google scholar IDs for each university (e.g., MIT's ID is 16345133980181568013). We next collect research profiles from each university, by scraping the URL associated with that university (e.g.,  https://scholar.google.com/citations? view\_op=view\_org \&org= 16345133980181568013 \&hl=en\&oi=io). Finally, we extract the name, affiliation, citations, and research topics of each individual researcher at that university, using a regular expression to match the relevant fields  from the html file. 

\medskip\noindent{\bf Data Cleaning:} 
In the early days, researchers creating google scholar profiles manually created their own research topics. This might account for the large number of typos and acronyms in the dataset. These days google auto-suggests relevant topics and allows up to five research topics per profile. 
We use comma as the primary topics separator and a regular expression to replace other separators (e.g., ... / ; .  \#) with a comma. For html tags we use beautifulsoup~\cite{bsoup}, a python package for cleaning up html tags. 

\medskip\noindent{\bf Topic Splitting and Merging: } 
Once the data above is collected and analyzed, it is easy to see that many topics should be split; see Table~\ref{split-label}. 
We split topics by pattern-matching conjunctions (i.e., or, and). 
\begin{table}[]
	\centering
	\begin{tabularx}{1\textwidth}{|X|}
		\hline
		\dots methods for longitudinal \textbf{or} clustered data, 
		statistics \textbf{for} neuroscience, \dots\\ \hline
		\dots data \textbf{and} model management, data mining, bioinformatics, algorithms \dots                        \\ \hline
	\dots new energy materials, Supercapacitor, photo \textbf{and} electro-catalysis of water \dots                                  \\ \hline
		\dots Thyroid, Nuclear Cardiology \textbf{and} Neurology, Gluconeogenesis,  \dots          \\ \hline
	\end{tabularx}
		\medskip\caption{Examples of records listing multiple topics that should be split.	\label{split-label}
}\vspace{-.6cm}
\end{table}

Merging is appropriate for topics that are similar but listed slightly differently. For example, out of out of half a million researchers in our dataset, 100 list $algorithm$, 20  list $algorithmics$, and 1,087 list $algorithms$. To handle this problem we need to determine the main topic with which the other topics should be merged. We use snowball~\cite{porter2001snowball} 
to find the root word by applying stemming (which removes endings such as $-s$, $-ed$, $-ing$). 
In the example above, snowball converted \ $algorithm$, $algorithmics$, and  $algorithms$ to the stemmed word  $algorithm$, however, applying snowball may result in stems that are difficult to understand (e.g., $applied$ and $applications$ are converted to $appli$). With this in mind, we set the main topic to be the one with highest frequency among all the topics with the same stem. 

\medskip\noindent{\bf Topic Correction:}  
Topic splitting and merging does not resolve all topic issues, as further modifications might be needed due to leading and trailing spaces, lower and upper case letters, punctuation and control characters, and duplicate words. Other issues of this type include ``Human Computer Interaction" and ``Computer Human Interaction," which are really the same topic. We try to address such issues with Google's Openrefine~\cite{OpenRefine} fingerprint key collision method, which attempts to find alternate representations of the same topic~\cite{cavnar1994n,hjaltason2003index,elmagarmid2007duplicate}.

\begin{figure}[b]
	\vspace{-.2cm}\hfil \includegraphics[width=1\textwidth]{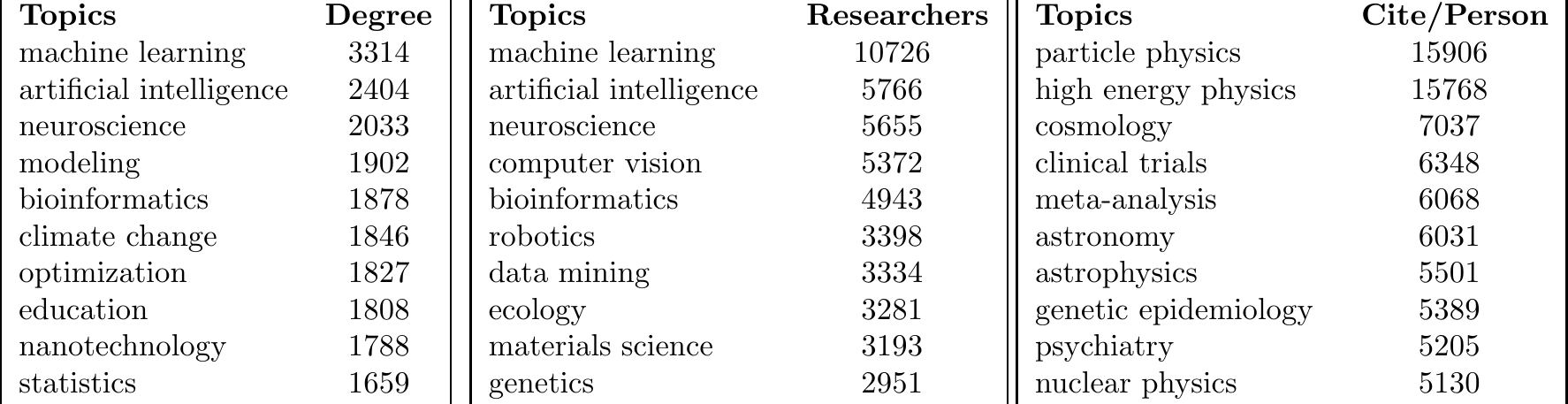} \hfil
	\vspace{-.3cm}\caption{Top ten topics: highest degree, number of researchers, number of citations.\vspace{-.5cm}}
	\label{fig:gstats}
\end{figure}

\begin{figure} 
	\hfil \includegraphics[width=0.7\textwidth]{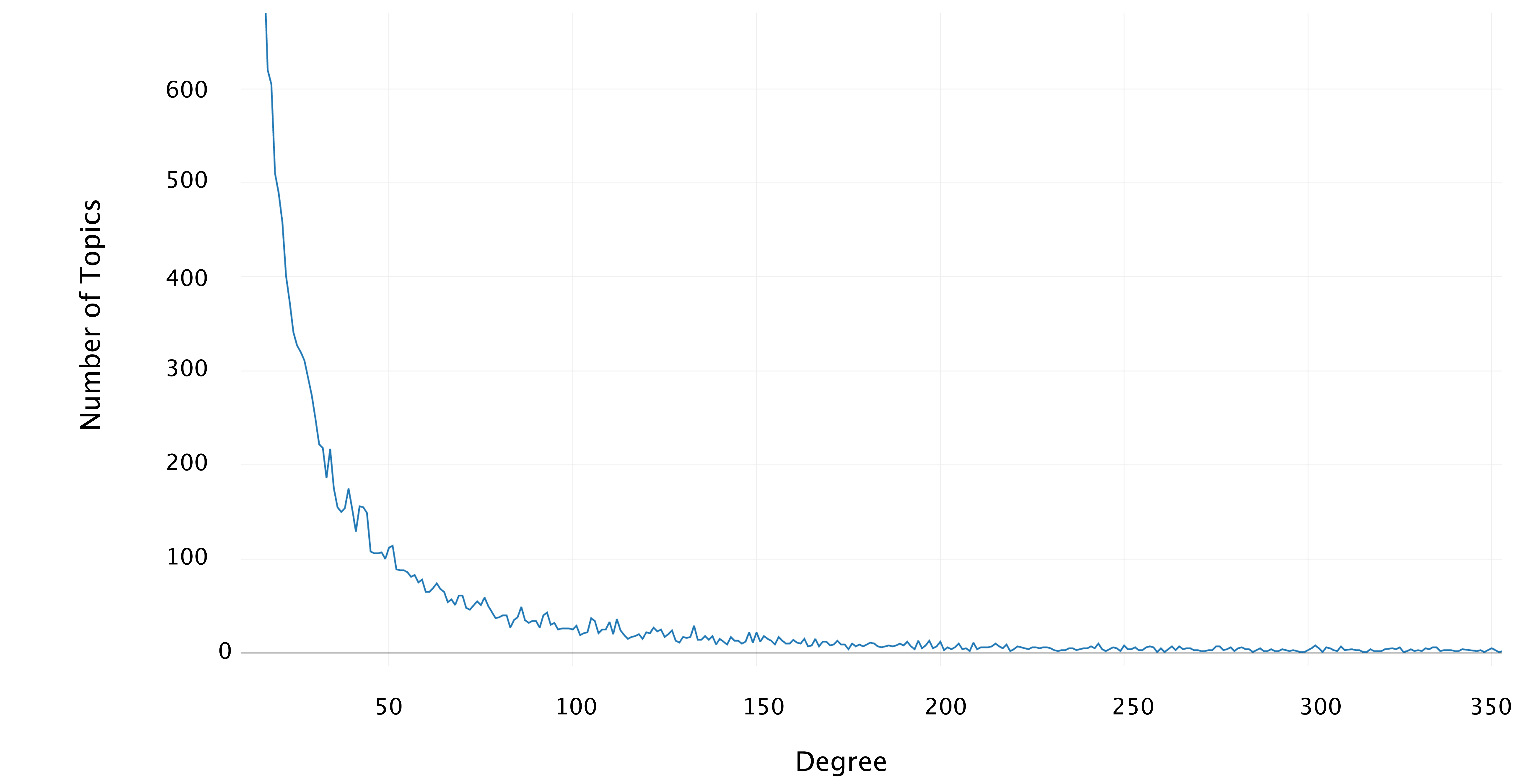} \hfil
	\caption{Degree distribution of the topics network.}
	\label{fig:nodegraph}
\end{figure}

\medskip\noindent{\bf Network Statistics:}
After the steps above, our topic graph contains 34,774 topics and 646,582 edges.  
There are 17  components, but just one giant connected component (34,741 nodes and  646,565 edges).
The average shortest path length is 3.141 which shows that the topic network is highly connected. The graph has a low global clustering coefficient of 0.09 (defined as the ratio of the number of triangles over the total number of node triples); see degree distribution in Fig.~\ref{fig:nodegraph}.
 The node ``machine learning" has the highest degree and more researchers are reporting working on this topic than any other. Figure~\ref{fig:gstats} shows the top ten topics by degree, by number of researchers, by citations per person. 

 Figure~\ref{fig:profiles} shows which institutions contributed the most scholar profiles. 
Interestingly, some universities seem to have more profiles than academic staff, (likely due to doctoral and postdoctoral students with university affiliations), although the majority of the universities are associated with fewer profiles than the size of their academic staff.

\vspace{-.2cm} 
\section{Map Generation}\vspace{-.3cm}
We use the GMap framework~\cite{Gansner10} to generate map layouts of the research topic graph and extend it to support semantic zooming. 
There are three high-level steps: (1) embedding the topic graph in the plane, (2) grouping vertices into clusters, and (3) creating the geographic map representation.  
We embed the graph using a scalable force-directed algorithm ({\tt sfdp} from graphviz) and then group the vertices using $k$-means clustering.
To create the geographic map look, we use a modified Voronoi diagram based on the obtained embedding and clustering. The geographic regions are colored such that no two adjacent countries have colors that are too similar, using the spectral vertex labeling method~\cite{Gansner10}.

\begin{figure}[t]
 	\hfil \includegraphics[width=1\textwidth]{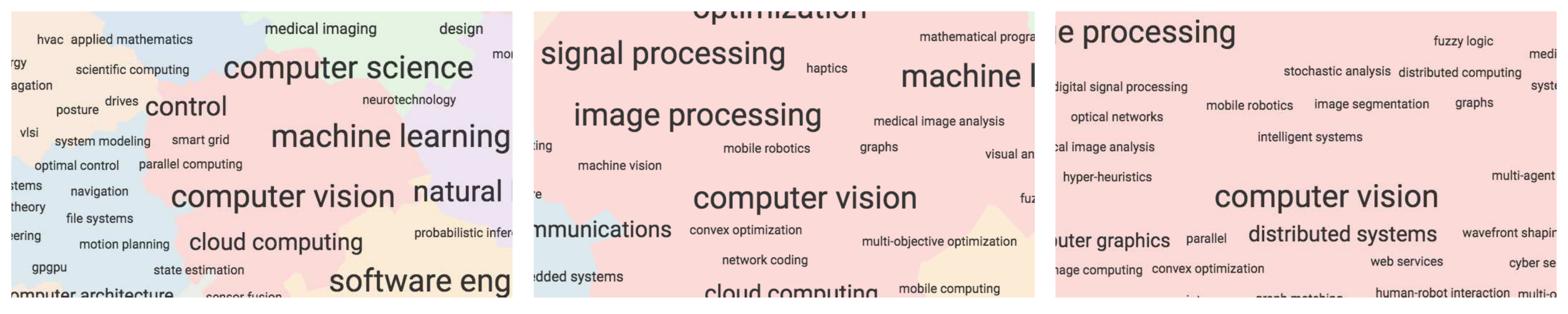} \hfil
 	\vspace{-.4cm}\caption{Three zoom-level views near the ``computer vision" topic.}\vspace{-.4cm}
 	\label{fig:zoom}
 	\end{figure}

We use the GraphViz implementation of node-overlap removal provided by {\tt prism}, but that provides non-overlapping labels only for the complete basemap, and not for the other 7 level-of-detail views, needed for semantic zooming.  Semantic zoom requires modifications to nodes, edges, clusters, and heatmaps. The google maps API handles all of these issues except for node-overlap (and hence node-label overlap), which is a natural side effect of zooming-in. To ensure that neither nodes nor labels overlap on any zoom-level, we compute different {\it node visibilities} for different zoom-levels. For each level, we sort the nodes by their weight (recall, that node weight is proportional to the number of researchers working on the topic associated with the node).  We make $i$-th node visible on the $j$-th level
if the bounding box of the $i$-th node does not overlap with the bounding boxes of  nodes $1,2,\cdots, (i-1)$. This algorithm takes $O(n^2)$ time but  it can be improved  by using~\cite{dwyer2005fast}.
Figure~\ref{fig:zoom} shows how the local neighborhood of ``computer vision" is changing in different zoom levels.

The size of the font label for topic $t$ is directly proportional to the number of researchers working on that topic, denoted by the weight: $w(t)$.
We assign font size from the range 80\% to 200\% of the default browser font size, as follows:

\begin{equation*}
\mathcal{F}_t =
\begin{cases}
80 & \text{if $w_t/10 \le 80  $} \\
200 &  \text{if $w_t/10 \ge 200 $} \\
w_t/10 & \text{otherwise}
\end{cases}
\end{equation*}

\begin{figure}
	\hfil \includegraphics[width=1\textwidth]{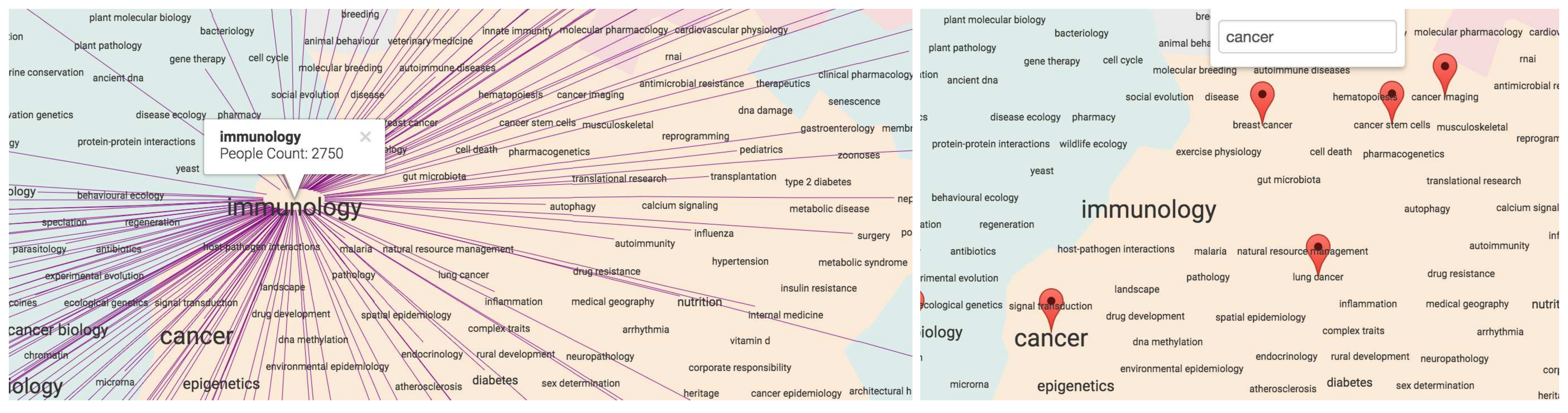} \hfil
	\vspace{-.5cm}\caption{Topic selection and search in {\tt rtopmap}.}
	\label{fig:nodeselection}
\end{figure}

\paragraph{\bf Web Interface and User Interaction:}
GMap produces a ``basemap"  from the given graph which is a static image that is not ideal for user interaction, such as zooming, panning, and searching. We enable interactions with the basemap with the help of the google maps API~\cite{svennerberg2010beginning}. Specifically, we take the output from GMap and convert it into google map objects, i.e., $google.maps.SymbolPath$, $google.maps.Polygon$, $google.maps.Polyline$, etc. 
For the web interface we provide 8 levels of details, showing different subgraphs, depending on the zoom level. 
We provide basic search functionality, which finds topics containing the query words. Clicking on a node shows the number of people who work on that topic in the underlying dataset and highlights edges to adjacent nodes (other topics that are frequently co-listed with that topic); see Fig.~\ref{fig:nodeselection}.
%
 \paragraph{\bf Basemaps:}
 We compute several different basemaps, covering a different set of universities. The full map is determined by researchers in 1,000 universities around the world~\cite{top1000}, but we also provide basemaps for universities in the United States and universities in Europe.
 Changing the basemap results in different node-weights and hence different label font-sizes. This is a useful feature when comparing a specific US university with universities around the world, with universities in the US, or with universities in Europe. 

 \paragraph{\bf Overlays:}
 After creating the basemap and all level-of-detail maps, we use overlays to show additional information, such as human resource investments of and citations associated with a specific university. Overlays can also be used to highlight topics associated with different departments (e.g., Computer Science, History) and even individual text documents (e.g., research paper, call for proposals). The overlay requests are collected from the browser (client) and the request is processed on the server, which then returns the necessary data to produce the overlay in the client. This is discussed in more detail in the next section.
 
\section{Knowledge Strengths and Weaknesses}
\begin{figure}[t]
	\hfil \includegraphics[width=1\textwidth]{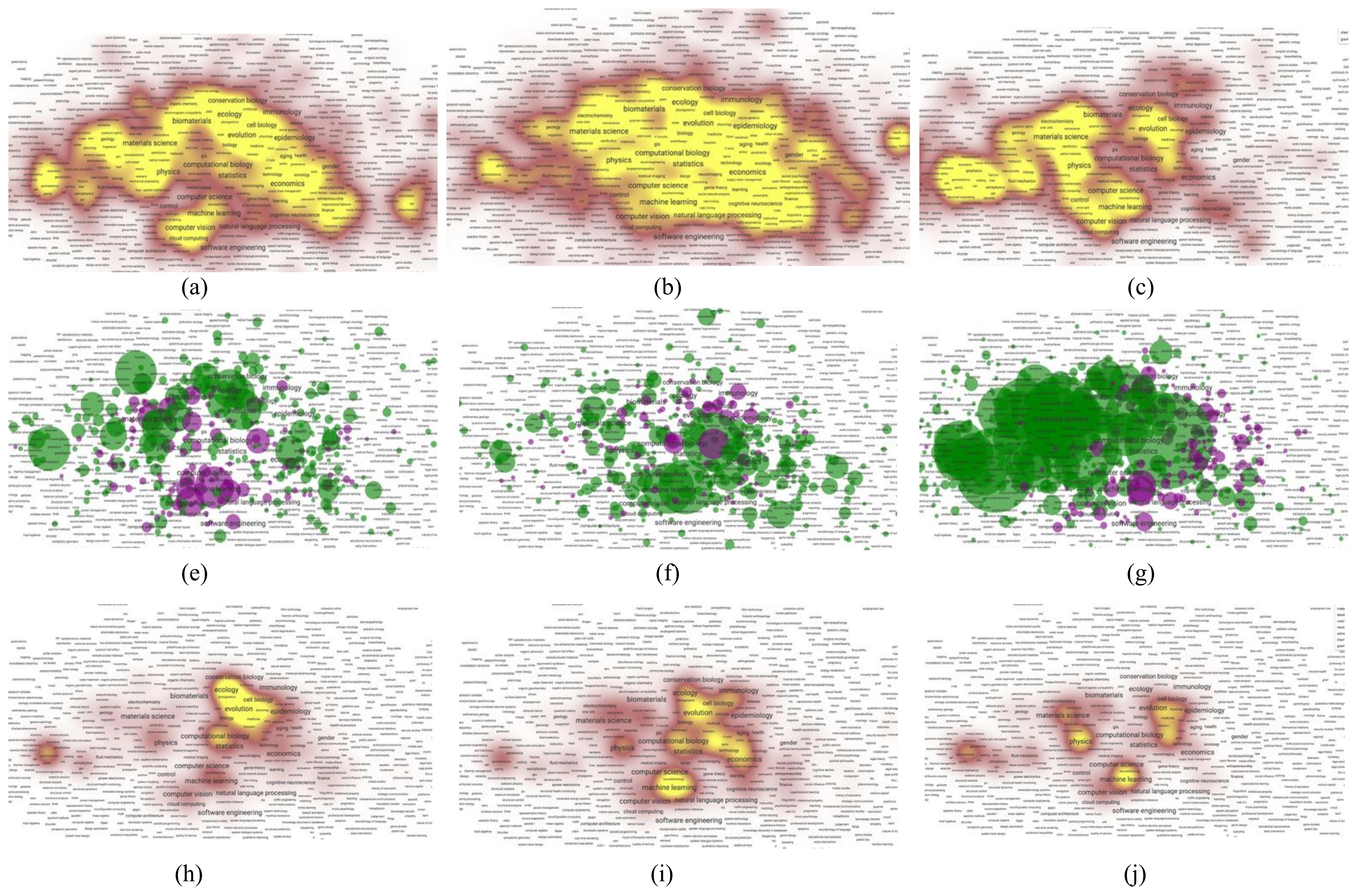} \hfil
	\vspace{-.3cm}\caption{
	University of Arizona (left column), Arizona State University (middle column), and California Institute of Technology (right column). (a-c) Heatmap overlays, based on citations. (e-g) Showing the number of people associated with different topics: green (purple) circles represent higher (lower) than average number of people working in this area. (h-j) Normalized citation heatmaps.  Higher resolution  images can be found in the image gallery at \url{rtopmap.arl.arizona.edu}.\vspace{-.2cm} }
	\label{fig:strengthandweakness}
\end{figure}

\vspace{-.3cm}
Quantifying knowledge strengths and weaknesses is a non-trivial challenge. We provide a university-level search feature which allows a specific university to be selected. We then attempt to visualize the strength and weakness of that university by computing the number of people working on different research topics and the number of citations associated with different topics for researchers from that university. Figure~\ref{fig:strengthandweakness} illustrates this using the University of Arizona (UofA), Arizona State University (ASU), and the California Institute of Technology (CalTech).  It is easy to see that UofA has a significant human resource investment in ecology/evolution (large green circles around these topics) and this translates to many citations in these topics. Such visualizations also make it possible to see that UofA has
not invested human resources in computer science (purple circles around CS topics) but CS is still associated with a large number of citations.

\vspace{-.2cm} 
\subsubsection{Citation Overlays: } 
Let $r$ be a researcher and let $topics(r)$ be the set of topics associated with researcher $r$. We denote number of citation received by $r$ as $cite(r)$.  
Then we can define $T$ as the set of topics that associated with university $X$ as follows:    
$T=\bigcup\limits_{\forall r \in X} topics(r)$. Then the citations associated with university $X$ for each each topic $t \in T$ is determined by the sum of citations of researchers at university $X$ who work on topic $t$: 
$c_X(t)=\sum\limits_{r\in X \& t\in r} cite(r)$.


The above formulas produce raw citation counts, although not all research fields cite at the same rate, e.g., ``particle physics" is associated with more citations than average (due to high number of co-authors and citations per paper); see also the citations-per-person table in Fig.~\ref{fig:gstats}. 
With this in mind we provide the option to normalize citation counts by total number of citations associated with a specific field $t$:  $normalized\ citation\ of\ t= c_X(t) \frac{c(t)}{C}$, where  $c(t)=\sum\limits_{r\in X \& t\in r} cite(r)$ and $C$ is the total number of citations for all topics.

When a researcher lists multiple topics, it is not easy to determine which citation contributes to which topic. In this case each citation contributes equally to each topic. A more careful analysis of the citation meta-data might allow us to distribute the contribution of each citation to different topics.

\subsubsection{Human Resources Overlays:}
We calculate human resource investment of a particular university by simply counting researchers and comparing the results to the averages. That is, to determine the human resource investment in topic $t$ at university X, given a base set (top 1,000 universities, US universities, European universities), we calculate the difference in the percentage of researchers at university $X$ who work on topic $t$ and the percentage of researchers in the base set who work on topic $t$. If this difference is positive (negative) then we consider this a human resource strength (weakness) of university X. This is illustrated with circles of different color: green for strength and purple for weakness. The size of the circles is proportional to the magnitude of the difference.
\begin{figure} 
	\vspace{-.3cm}	\hfil \includegraphics[width=1\textwidth]{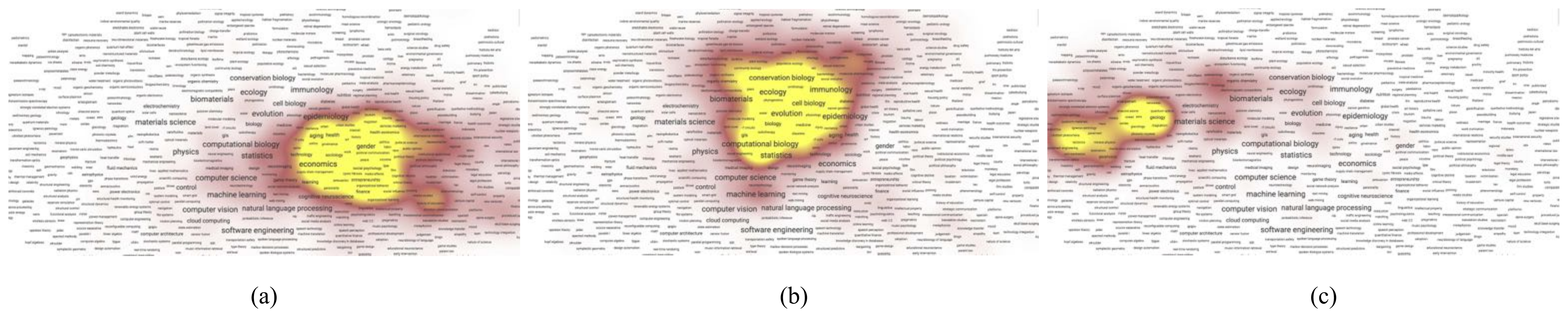} \hfil
\vspace{-.4cm}	\caption{Department visualization on the topic map. (a) Department of Economics, (b) Department of Biology and (c) Department of Geology.}\vspace{-.4cm}
	\label{fig:dept}
\end{figure}
\vspace{-.4cm} 

 \paragraph{\bf Department Overlays:}
We can also visualize what part of the map is associated with what academic department using map overlays. Figure~\ref{fig:dept} shows three such department heatmap overlays. 
Different universities give different names to the same department, or group together academic units in colleges and schools. 
We match a given a keyword, such as economics, biology, geology, with the affiliation-information in our database. Google scholar affiliations are a combination of designation, department name, university name, as shown in Table~\ref{table:dept}.  We use a regular expression to match researchers based on department name, which gives us the topics associated with the department, and eventually the weights, which are used to build the heatmap.
 
\begin{table}[]
	\centering
	\vspace{-.4cm}\begin{tabularx}{1\textwidth}{|X|}
		\hline
 					 {Affiliation}  \\\hline
Professor of \textbf {Computer} Science and Artificial Intelligence, Granada University\\
Professor of  \textbf{Chemistry} and Chemical \textbf{Biology}\\
\textbf{Computer} Science, Virginia Commonwealth University\\
Professor of   \textbf{Biochemistry} and  \textbf{Molecular} Biophysics, Washington University\\\hline

	\end{tabularx}
\medskip\caption{Examples of affiliations in google scholar.}\vspace{-.7cm}
\label{table:dept}
 	\end{table}

\begin{figure}
	\vspace{-.3cm}\hfil \includegraphics[width=1\textwidth]{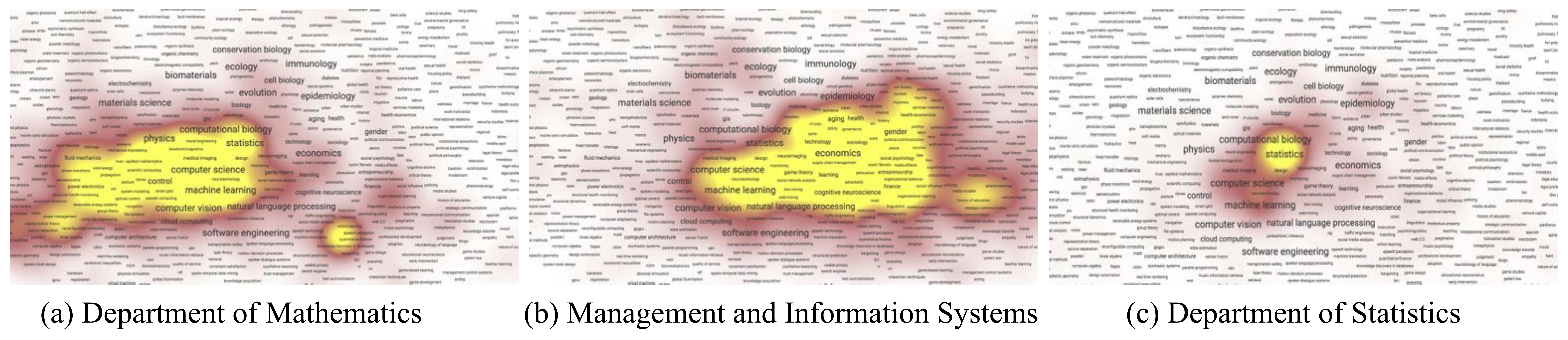} \hfil
\vspace{-.3cm}\caption{ Topics overlapped with different departments and institutes.}\vspace{-.5cm}
	\label{fig:topicsoverlap}
\end{figure}

 Using such visualizations, we can find research fields that are shared by different departments, e.g.,  ``machine learning," which is studied in mathematics, management and information systems, and statistics; see Fig.~\ref{fig:topicsoverlap}.

\begin{figure} 
	\hfil \includegraphics[width=1\textwidth]{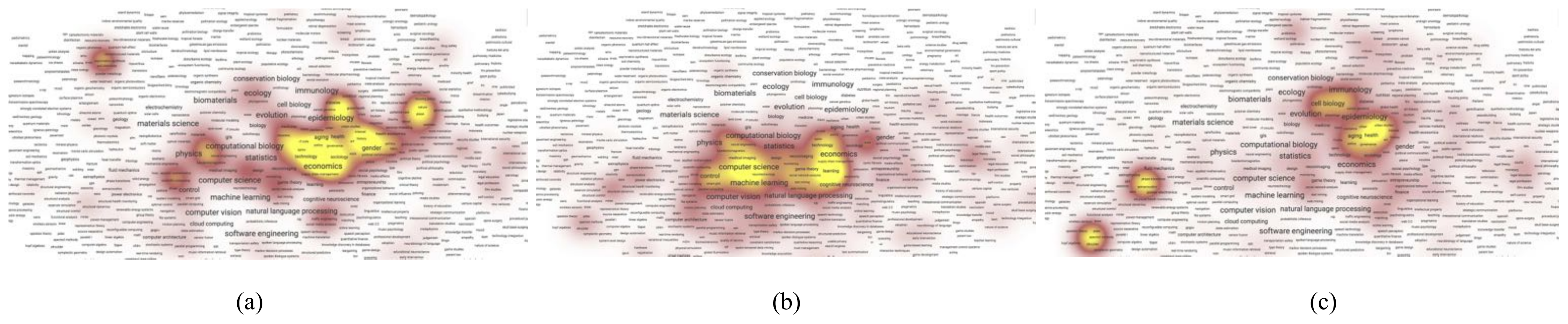} \hfil
\vspace{-.5cm}	\caption{Documents: (a) research paper on economics~\cite{Motesharrei201490},  (b) call for paper for CS conference NIPS~\cite{nips2017}, and (c) call for proposals for National Institute of Health~\cite{nih}.}\vspace{-.7cm}
	\label{fig:docoverlay}
\end{figure}

 \paragraph{\bf Document Overlays:}
We can also visualize documents as map overlays.  We extract research terms and their frequency from a given input document, usually a URL. This requires a cleanup of the text, tokenization, and stemming. 
We next compute the term frequency of unigrams, bigrams, and trigrams, which become our \textit{candidate research topics}. Then we match the candidate research topics with the research topics that are already in our database. The result is a collection of research topics associated with the document along with weights, which are used to create the heatmap overlay; see 
  Fig.~\ref{fig:docoverlay}.

%
%

\section{Implementation}\vspace{-.3cm}
We use a variety of tools to process, clean, store, and process our data: mongodb scripts, sqllite, python, R, Java-Lucene, openrefine. Google maps API and jquery are used for map drawing and to handle user interaction in the web application. We run python-django for the webserver and mongodb for database storage and query.
 

For each researcher, our database stores name, google scholar id, university id, total citations, domain name of email address, affiliation, raw research areas, research phrases, and stemmed phrases.   

The default settings for filtering the network (removing nodes and edges with low weights) results in the more compact version of the network with 6,052 nodes and 26,162 edges.  Generating the topics map in svg format (layout, clustering, node-overlap removal, etc.) takes 14 seconds.
Loading the initial base map takes 12,638ms,  including  7,836ms for scripts,  2,444ms for rendering, and 275ms for painting (in google chrome v. 58 browser). 
 Hiding the edges results in much faster interaction and this is indeed the default setting.
 Interaction with the basemap, map navigation, zooming with edges takes 1,515ms. 
 Computing human resource overlays takes  3,129ms, while citation heatmaps require 1,231ms. 

The {\tt rtopmap}  system currently runs as a virtual machine on a Dell PowerEdge R430 server with 2 Intel(R) Xeon(R) CPU E5-2530 v4 @2.20GHz processors and 32GB of memory.


\section {Discussion and Limitations}


We use google scholar as the source for our data, with all of the advantages (e.g., a great deal of information) and disadvantages (e.g., the data is not curated) that this choice is associated with.  Further, different research areas differ in their representation on google scholar. For example, there seem to be many more computer science and physics profiles than history and psychology ones.  Researchers from different universities also use google scholar profiles at different rates. 

Once the data has been gathered, the set of universities used to create the basemap has a non-trivial impact. Focusing only on English language terms, also biases the results. Despite our attempts to clean, split and merge topics, many issues remain. For example, researchers sometimes use acronyms (NLP for natural language processing, or HCI for human computer interaction) which should be expanded and merged.

Our strengths and weaknesses calculation, based on human resource investments, is associated with other biases: the human resource counts (from Wikipedia) are not guaranteed to be accurate, we do not distinguish tenure-track faculty from other type of staff (e.g., doctoral and postdoctoral students). 
Our citation-based calculations are also biased and inaccurate, as google scholar often misattributes papers and we cannot match specific citations to specific topics associated with a researcher, or distribute the citation contribution among its co-authors. 


 \vspace{-.3cm} 

	\section{Conclusions and Future Work}\vspace{-.3cm}
We presented {\tt rtopmap}: a system for visualizing and analyzing research topics. Despite non-trivial limitations, we believe that the system is useful as it gathers this information in self-reported, bottom-up fashion, rather than the more traditional top-down hierarchical taxonomies and ontologies. With the help of map overlays, {\tt rtopmap} makes it possible to visualize human resource investments and scholarly output for different academic institutions. Department profiles and documents can also be visualized via overlays. Finally, the system implements in-the-browser, map-based interactive navigation of the fairly large underlying network, supporting panning, zooming, and searching. The {\tt rtopmap} system is available at \url{http://rtopmap.arl.arizona.edu/.}

A natural extension of this work would be to match calls for proposals with individual researchers or groups of researchers at a specific university, making it possible to quickly identify potential participants in multi-disciplinary research proposals. Adding more data (e.g., funding statistics from national funding agencies, patents, media coverage of research projects) can augment the picture of a specific university, or enable more detailed comparisons between different universities. The visualization system itself is in a prototype stage. Our goal is to make it more responsive, improve HCI aspects, and extend its functionality to smaller screens.

	\subsubsection*{Acknowledgments}
	We thank Nirav Merchant, Mihai Surdeanu, and the Data7 Institute at the University of Arizona.   

\newpage
	\bibliographystyle{splncs03}
\bibliography{topicsmap}
\newpage
\section*{Appendix}

We place several additional tables and figures in this section.

\begin{figure}[h]
	\begin{center}
		\vspace{-.3cm}\begin{tabular}{|lcc|}
			\hline
			\centering 
			\textbf{Name of University} & 	\textbf{	Academic Staff }	 &	\textbf{\# Profiles}\\\hline
			
			Stanford University	&2118&	8104\\ 
			University of Washington&	5803&	5562\\ 
			Harvard University&	4671&	5356\\ 
			Massachusetts Institute of Technology&	1021&	3527\\ 
			University of Michigan&	6771&	3413\\ 
			University of Toronto&	2547&	3148\\ 
			University of Cambridge&	6645&	2669\\ 
			Texas A\&M University&	2700&	2515\\ 
			University of Minnesota&	3804&	2511\\ 
			Pennsylvania State University&	8864&	2368\\\hline
		\end{tabular}
		\vspace{-.5cm}\caption{Academic staff numbers according to Wikipedia entries.\vspace{-1cm}}
		\label{fig:profiles}
	\end{center}
\end{figure}
\vspace{-.5cm}

\vspace{-.8cm}

\vspace{-.5cm}

\end{document}